\documentclass[journal]{IEEEtran}
\usepackage[dvipsnames]{xcolor}

\usepackage{graphicx,url}
\usepackage[american]{babel} 
\usepackage{soul}
\usepackage{amssymb}	 
\usepackage{acronym}
\usepackage{amsfonts}
\usepackage{multirow} 
\usepackage{longtable}
\usepackage{paralist}
\usepackage{float}
\usepackage{setspace}
\usepackage{xcolor}
\usepackage{enumerate}
\usepackage{flushend}
\usepackage{tabularx}
\usepackage{paralist}
\usepackage[normalem]{ulem}
\usepackage{amsmath}
\usepackage{multirow}
\usepackage{multicol}
\usepackage[utf8]{inputenc}
\usepackage{subfig} 
\usepackage{subcaption}
\usepackage{setspace}
\usepackage{url}
\usepackage{tikz}
\newcommand*\numrounded[1]{\tikz[baseline=(char.base)]{
            \node[shape=circle,draw,inner sep=0.7pt] (char) {#1};}}

\usepackage{enumitem}
\usepackage{algorithm}
\usepackage[noend]{algpseudocode}

\usepackage{algorithm, algpseudocode, times}

\newcommand{\final}[1]{{\color{black}#1}}
\usepackage[normalem]{ulem} 

\setul{0.15ex}{0.8pt}         

\newcommand{\eg}{\textit{e.g.,} } 
\newcommand{\ie}{\textit{i.e.,} } 

\ifCLASSINFOpdf
  
\else
  
\fi

\begin{document}

\title{eXplainable Artificial Intelligence for RL-based Networking Solutions}

\author{\IEEEauthorblockN{Yeison S. Murcia, Oscar M. Caicedo, Daniela M. Casas, and Nelson L. S. da Fonseca\\}
        
 	    
\thanks{This study was partly financed by the Brazilian Council for Scientific and Technological Development (CNPq) grant \# 427814/2018-9, CNPq grant \# 405940/2022-0 and CAPES grant \# 88887.954253/2024-00, and the S\~{a}o Paulo Research Foundation (FAPESP), Brazil, under Grant 2023/00673-7.}
\thanks{Yeison S. Murcia, Daniela M. Casas, and Nelson L. S. da Fonseca (corresponding author) are with the University of Campinas (UNICAMP), Brazil. E-mail: \{yeisonm, danielac\}@lrc.ic.unicamp, nfonseca@ic.unicamp.br.}
\thanks{O. M. Caicedo is with the Departamento de Telematica, Universidad del Cauca, Popay\'an, Colombia. E-mail: omcaicedo@unicauca.edu.co.}
}

\maketitle

\acrodef{CNN}{Convolutional Neural Networks}
\acrodef{CP}{Control Plane}
\acrodef{DNN}{Deep Neural Networks}
\acrodef{DP}{Data Plane}
\acrodef{DT}{Decision Tree}
\acrodef{MP}{Management Plane}
\acrodef{KP}{Knowledge Plane}
\acrodef{AI}{Artificial Intelligence}
\acrodef{API}{Application Program Interface}
\acrodef{BGP}{Border Gateway Protocol}
\acrodef{CSV}{Comma-separated values}  
\acrodef{DL}{Deep Learning}
\acrodef{DRL}{Deep Reinforcement Learning}
\acrodef{DQN}{Deep Q-Network}
\acrodef{DQL}{Deep Q-learning}
\acrodef{CRP}{Cognitive Routing Plane}
\acrodef{EWBI}{East/Westbound Interface}
\acrodef{FM}{Full Mesh}
\acrodef{IoV}{Internet of Vehicles}
\acrodef{ICE}{Individual Conditional Expectation}
\acrodef{IoT}{Internet of Things}
\acrodef{KDN}{Knowledge-Defined Networking}
\acrodef{LIME}{Local Interpretable Model-Agnostic Explanations}
\acrodef{LLDP}{Link Layer Discovery Protocol}
\acrodef{LL}{Logical Links}
\acrodef{MDP}{Markov Decision Process}
\acrodef{MI}{Management Interface}
\acrodef{ML}{Machine Learning}
\acrodef{MSE}{Mean Square Error}
\acrodef{NBI}{Northbound Interface}
\acrodef{NN}{Neural Network}
\acrodef{NOS}{Network Operative System}
\acrodef{PDP}{Partial Dependence Plot}
\acrodef{PL}{Physical Links}
\acrodef{PPO}{Proximal Policy Optimization}
\acrodef{OSPF}{Open Shortest Path First}
\acrodef{QoS}{Quality of Service}
\acrodef{RIP}{Routing Information Protocol}
\acrodef{RL}{Reinforcement Learning}
\acrodef{RCM}{Root Control Module}
\acrodef{RSIR}{Reinforcement Learning and Software-Defined Networking Intelligent Routing}
\acrodef{RRM}{Radio Resource Management}
\acrodef{SBI}{Southbound Interface}
\acrodef{SDN}{Software-Defined Network}
\acrodef{SD-WAN}{Software-Defined Wide Area Network}
\acrodef{SHAP}{Shapley Additive Explanations}
\acrodef{SL}{Supervised Learning}
\acrodef{SLA}{Service Level Agreements}
\acrodef{TCP}{Transmission Control Protocol}
\acrodef{TM}{Traffic Matrices}
\acrodef{VM}{Virtual Machine}
\acrodef{WAN}{Wide Area Network}
\acrodef{UAV}{Unmanned Aerial Vehicles}
\acrodef{USL}{Unsupervised Learning}
\acrodef{VR}{Virtual Reality}
\acrodef{VM}{Virtual Machine}
\acrodef{WSN}{Wireless Sensor Network}
\acrodef{UDP}{User Datagram Protocol}
\acrodef{XAI}{Explainable Artificial Intelligence}
\acrodef{XDRL}{Explainable Deep Reinforcement Learning}
\acrodef{XRSIR}{Explainable RSIR}
\acrodef{XRL}{Explainable Reinforcement Learning}
\acrodef{ZC}{Zone Controller}

\begin{abstract}


Reinforcement Learning (RL) agents have been widely used to improve networking tasks. 
However, understanding the decisions made by these agents is essential for their broader adoption in networking and network management. To address this, we introduce eXplaNet—a pipeline grounded in explainable artificial intelligence —designed to help networking researchers and practitioners gain deeper insights into the decision-making processes of RL-based solutions.
We demonstrate how eXplaNet can be applied to refine a routing solution powered by a Q-learning agent, specifically by improving its reward function. In addition, we discuss the opportunities and challenges of incorporating explainability into RL to optimize network performance better.

\end{abstract}

\IEEEpeerreviewmaketitle

\section{Introduction}
\IEEEPARstart {T}{he} rapid proliferation of technologies and services such as the Internet of Things, Unmanned Aerial Vehicles, and Virtual Reality, combined with the accelerated development of mobile networks, 
is generating vast volumes of heterogeneous and complex data.
\ac{ML} techniques, particularly \ac{DL} and \ac{RL}, have become fundamental tools for solving data deluge-generated critical network issues to enhance adaptability and operational efficiency in network management \cite{1}. 
However, the real-world deployment of ML models in communication networks remains limited. This is due primarily to the lack of explainability, interpretability, and transparency, which fosters skepticism and distrust among network operators \cite{9crack}, \cite{10trus}.

\ac{XAI} addresses these concerns by providing transparency and interpretability to ML models, which are often treated as “black boxes” \cite{xai6g}. In both traditional and modern \acp{SDN}, \ac{XAI} strengthens trust by enabling operators to understand the rationale behind agents' decisions. This understanding is beneficial for debugging and fine-tuning tasks such as resource allocation and routing policy optimization. Specifically, \ac{XRL} provides insights into agents' reasoning, policies, and actions, allowing network administrators to identify biases or failures in real time and correct them before the network's performance degrades \cite{12xrl1}. 

\final{Networks are inherently dynamic, high-dimensional, and constrained by latency, privacy, and reproducibility requirements. Applying \ac{XAI} directly to the original RL agent in networks often demands extensive explorations and evaluations, which leads to substantial computational complexity and costs \cite{1} \cite{metis}. To address this challenge, surrogate models have been adopted in RL-governed networks. These models approximate the agent’s policy and enable interpretation through feature relevance analysis \cite{10trus} \cite{surrogateredes} \cite{9829396} \cite{XRLredes2}, achieving a balance between fidelity to the original agent and computational efficiency. Despite these advances, an important gap remains: the knowledge obtained from explanations is seldom leveraged to improve the agent’s policy itself. While \ac{XAI} has proven effective for understanding and validating agent decisions, its explanatory insights are rarely transformed into actionable strategies for policy optimization. Explanations generated by surrogate models and feature relevance analysis usually remain external to the agent’s learning loop, rather than feeding back into the policy to refine behavior and performance.}

\final{To address the gap mentioned above, we introduce an XAI-based pipeline called eXplaNet, which systematically transfers explanatory knowledge into policy optimization, prioritizing the most relevant information and guiding agent adjustments to enhance network performance. The pipeline begins by defining a dataset that records network states and agent actions. Various \ac{XAI} surrogate models are then trained to approximate the policy, and the one that best mimics the original RL model is selected. \ac{XAI}-based feature relevance analysis is then applied to this surrogate. The extracted knowledge is subsequently integrated into the agent, guiding policy refinement and prioritization of impactful features. Through this process, eXplaNet delivers not only greater transparency but also measurable performance improvements and enhanced operational confidence.}

Additional original contributions in this paper are: \numrounded{1} A showcase of eXplaNet via \ac{XRSIR}. \ac{XRSIR} improves \ac{RSIR} \cite{6rsir}, our \ac{RL}-based SDN routing solution. 
\ac{XRSIR} improves both the explainability and performance of RSIR. 
Results show that eXplaNet is a practical approach for enhancing RL
through explanations, bringing RL closer to real-world applicability in
networking. \numrounded{2} A comprehensive discussion of lessons learned, research opportunities, and the key challenges involved in integrating \ac{XAI} with \ac{RL} in the context of networking.

This paper is organized as follows. Section II addresses the use of \ac{XAI} in networking. Section III introduces eXplaNet. Section IV presents \ac{XRSIR}, a routing solution derived by employing eXplaNet. Section V suggests future research directions, and Section VI draws some conclusions.

\section{\ac{XAI} for Reinforcement Learning }
\subsection{Reinforcement Learning}
\ac{RL} is a powerful paradigm within \ac{ML} for solving decision-making problems through an automated trial-and-error process. An \ac{RL} agent periodically interacts with an environment and receives feedback in the form of rewards that reflect the effectiveness of those actions. \ac{RL} problems are typically modeled as a Markov Decision Process, comprising three key components:
a state space \(\mathit{S}\), an action space \(\mathit{A}\), and an immediate reward function \(\mathit{R}(s_t, a_t, s_{t+1})\) \cite{42drl}. The primary objective of an \ac{RL} algorithm is to determine an optimal policy for maximizing long-term rewards in an environment, considering available states and candidate actions per state.

\subsection{\ac{XAI} techniques for networking}

The rapid evolution of networks and the increasing demand for services have led to the proliferation of AI-driven networking solutions. In this context, \ac{RL} stands out as an attractive solution for optimizing resource management and routing policies, among others, since it offers more adaptive and efficient responses to the changing needs of modern networks. Despite its potential to dynamically adapt to different topologies and network demands, adopting \ac{RL} in production environments faces multiple challenges \cite{9crack} \cite{metis} related to interpretability, manageability, and reliability.

A significant limitation of \ac{RL}-based approaches is their lack of transparency—network operators often cannot discern why an agent selects a particular policy, especially during abnormal conditions such as link failures or traffic spikes. This obscurity hinders troubleshooting and fine-tuning. Manageability becomes problematic in large-scale networks when the training process fails to highlight the most influential features or variables. Moreover, the inability to audit the decision-making process can undermine trust and reliability, especially in critical applications such as \ac{SDN} supporting ultra-reliable low-latency communication.

\ac{XAI} addresses these challenges by introducing transparency into the decision-making process of \ac{RL} agents. By enabling the visualization and interpretation of internal mechanisms, \ac{XAI} facilitates the identification of key decision drivers, supports model debugging, and guides performance optimization. These capabilities are crucial for enhancing safety, reliability, and network practitioners trust while minimizing the risk of adversarial exploitation. Consequently, \ac{XAI} plays a vital role in encouraging the broader deployment of \ac{RL} in production network environments \cite {1}.

\ac{XAI} techniques can be classified by their scope of interpretation. Global methods aim to explain the overall model's behavior, while local methods explain individual predictions \cite{xai6g}. Techniques such as surrogate models, \ac{SHAP}, \ac{LIME}, \ac{ICE}, and \ac{PDP} allow visualizing internal model processes, analyzing feature relevance, and simplifying models to create more explainable, interpretable, and trustworthy ones \cite{xai4.0}.

\final{Surrogate models are post-hoc \ac{XAI} methods that approximate either a black-box model or an \ac{RL} policy by using an interpretable proxy model (\eg decision trees or sparse linear models). They provide both \textit{local} and \textit{global} explanations. Local explanations are possible since surrogate models approximate the Q-values for pairs of states/actions, while global explanations are feasible due to the training of surrogate models using data that mimic the overall policy. 
To verify if the proxy model faithfully represents the original agent, fidelity metrics can be used.}

\final{\ac{SHAP} offers global explainability by applying cooperative game theory to assign each feature an importance value based on its average marginal contribution across all feature combinations, providing a principled measure of relevance. It also supports local explanations and captures complex feature interactions, enabling a detailed understanding of model behavior \cite{xai6g} \cite{XRLredes2}. In contrast, \ac{LIME} emphasizes local interpretability, approximating model behavior near a specific prediction with a simplified linear model. While effective for individual decisions, its global explanatory power is limited.
} 

\final{
\ac{PDP} and \ac{ICE} serve to complement \ac{SHAP} and \ac{LIME}. \ac{PDP} allows analyzing how the average model prediction changes when one or more features are systematically varied while holding others constant. This method produces visualizations that are particularly useful for identifying and interpreting global trends in the model’s decision-making process. \ac{ICE}, on the other hand, provides a more granular analysis by illustrating how changes in a feature affect individual instances. Unlike \ac{PDP}, \ac{ICE} captures variability across instances, making it especially valuable for instance-level investigations.
}

\final{Applying \ac{SHAP}, \ac{LIME}, \ac{PDP}, and \ac{ICE} to a surrogate model enables cross-validation of explanations and interpretability at local, global, and feature levels while assessing surrogate fidelity from multiple perspectives. \ac{SHAP} integrates global and local views via principled importance measures; \ac{LIME} focuses on individual predictions; \ac{PDP} highlights global feature effects; and \ac{ICE} exposes instance-specific behaviors often hidden in aggregates. Together, these methods provide a comprehensive framework for robust and multifaceted model interpretation, mitigating the risk of relying on a single technique.
}


\subsection{\ac{XRL} in networking: What has been done?}

\begin{figure*}[!htb]
    \centering
    \includegraphics[width=0.7\linewidth]{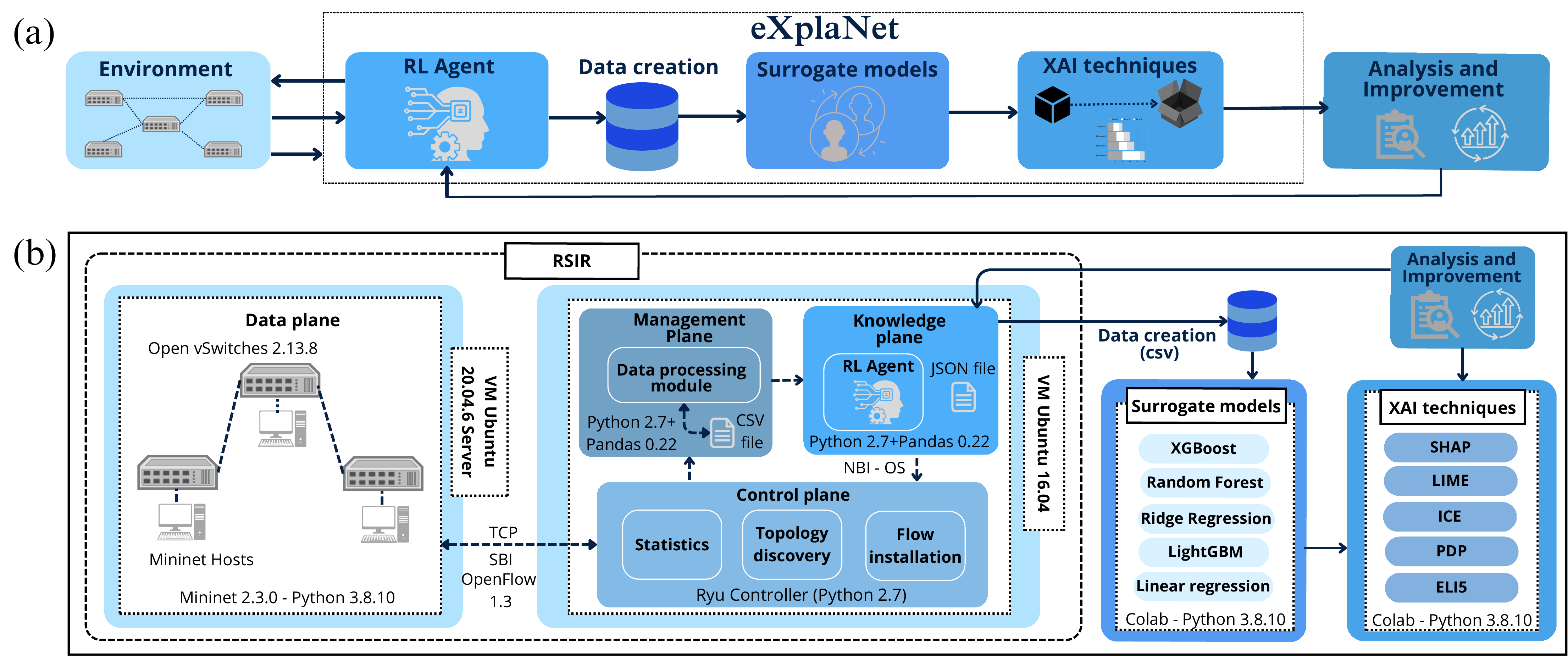}
    \caption{\final{Structure of the eXplaNet pipeline (a) and application to the RSIR case study (b) with interactions between systems and modules used.}}
    \label{fig:marco}
\end{figure*}

Recent research has enforced \ac{XAI} techniques into \ac{RL}-based networking solutions 
to enhance transparency, build confidence in decision-making, and facilitate model debugging and optimization.
The TRUSTEE framework \cite{10trus} uses decision trees to provide post-hoc explainability to \ac{ML} models in network security, including \ac{RL} ones. TRUSTEE 
enables network managers to comprehend the models' decision-making processes and assess whether they are affected by inductive biases. The work \cite{surrogateredes} proposes an \ac{XAI}-based framework to improve the trustworthiness of network management solutions aided by \ac{RL}. Similar to TRUSTEE, this framework employs decision trees as surrogate models to mimic and visually explain the decisions made by Q-learning agents, allowing network managers to grasp the rationale behind these decisions.

The work \cite{XRLredes2} combines \ac{SHAP} and an entropy mapper to quantify uncertainty in predictions, thereby improving trust and reliability in resource allocation in 6G networks, and seeking to meet Service Level Agreements transparently. The work \cite{metis} introduces the Metis framework, designed to provide interpretability to \ac{RL}-based network routing systems in \acp{SDN}. Metis uses hypergraphs to represent complex relationships and identify critical connections to optimize routes.
\section{eXplaNet}

Few investigations have explored the enforcement of \ac{XAI} into \ac{RL} models \cite{10trus} \cite{metis} \cite{surrogateredes} \cite{XRLredes2}, which hinders their deployment in real-world networks. Remarkably, those investigations fail to leverage explanations to improve the models. We developed eXplaNet based on the premise that the practical application of \ac{RL} in computer and telecommunications networks can benefit from incorporating explainability findings to improve \ac{RL} models. eXplaNet is a pipeline composed of several logical steps that facilitate the enhancement of \ac{RL} models using insights derived from surrogate models. \final{Our pipeline stands out for its inherent independence of the network topology. Its modular, data-driven design—linking network metrics with agent decisions—renders it agnostic to the underlying physical structure. Moreover, by incorporating model-agnostic \ac{XAI} techniques, such as surrogate models, \ac{SHAP}, \ac{PDP}, and \ac{ICE}, the pipeline ensures broad applicability across diverse learning models. As a result, it can be seamlessly transferred and adapted to optimize heterogeneous network solutions.}


\subsection{Overview}
Figure 1(a) introduces eXplaNet, a novel pipeline conceived, designed, and developed to explain and improve \ac{RL}-based networking solutions by \ac{XAI} enforcement. eXplaNet generates understandable explanations regarding the behavior of surrogate models that mimic \ac{RL} agents. These insights enable informed adjustments to the original \ac{RL} models, ultimately enhancing network performance. 

The eXplaNet's steps are Data Creation, Surrogate Model, \ac{XAI} Techniques, and Analysis and Improvement. The Data Creation step involves constructing a dataset using data collected from the original \ac{RL}-based solution while running the original model. The Surrogate Model step develops, trains, and evaluates various surrogate models and selects the model that best mimics the original one; a surrogate model offers an efficient, comprehensible, and flexible solution to explain the decisions made by a complex \ac{RL} agent \cite{surrouso4} since it allows understanding the behavior of the original model by leaving out all the intricate details of its inner workings \cite{surrouso}. 
The \ac{XAI} Techniques step involves applying various local and global \ac{XAI} methods to the selected surrogate model. 
The Analysis and Improvement step analyzes the results obtained from the previous step to gain detailed and accessible insights into the importance of features and enhance the understanding of the decision-making process of the \ac{RL} agent. Based on findings from the analysis, adjustments are made to the original model, such as removing features, modifying the reward function, and changing the exploration or exploitation method to improve its performance and RL-based network management.

\subsection{Steps}
\subsubsection{Data Creation} 
This step builds a dataset for training the surrogate model of the original \ac{RL} model. Data can be collected from various sources, such as values calculated and stored in the \ac{RL} agent's Q-table, which is generated through the Q-learning algorithm. Additionally, the dataset should include input features and the expected output values, which facilitate evaluating how these features influence decision-making. Preprocessing tasks are essential to ensure the quality and usefulness of the dataset. These tasks include data cleaning to remove outliers or inconsistent values, normalizing features to a comparable range, and encoding categorical variables to transform them into suitable representations for the model. These tasks ensure the dataset is consistent and ideal for efficiently training the surrogate model.


\subsubsection{Surrogate Model} This step focuses on building a surrogate for the \ac{RL} model. The proposed pipeline advocates using surrogate models because they operate based on a set of simple and understandable rules that mimic the behavior of \ac{RL} agents, thereby addressing their black-box nature. Furthermore, surrogate models favor the enforcement of \ac{XAI} techniques and are suitable for environments where it is infeasible to evaluate the outcome of interest directly. Creating a surrogate model involves selecting an inherently interpretable model (\eg a linear model or a decision tree) and training it with the dataset generated in the first pipeline step. Once trained, the surrogate model is evaluated to determine its fidelity in reflecting the predictions of the black-box model. Finally, the results obtained are analyzed. The surrogate model to select must be the one that best mimics the original model. 

\final{Given the inherent network dynamics, the surrogate model must be periodically updated to assimilate new observations. Depending on operational constraints and data availability, this adaptation can be realized by combining slide-window and experience replay with incremental training, selective retraining, or weighted learning or by parameter refinement via online calibration. Consequently, the surrogate model co-evolves with the network’s state, thereby preserving interpretability while maintaining a high fidelity to the underlying RL policy.}

\subsubsection{XAI Techniques} 
This step applies \ac{XAI} techniques to the surrogate model to understand how it makes decisions and, consequently, to explain the behavior of the original model. Key techniques in this regard are \ac{SHAP}, \ac{LIME}, \ac{PDP}, and \ac{ICE}, each offering different advantages and limitations. \ac{SHAP} plays a crucial role in assessing the global impact of features and their local interactions, while \ac{LIME} facilitates a detailed analysis of specific instances. \ac{PDP} and \ac{ICE} provide intuitive visualizations that enable the interpretation of feature influence on model predictions.

\subsubsection{Analysis and Improvement} This step aims to optimize the RL-based networking system. The results of the previous step are analyzed in depth, including graphical representations, explanatory diagrams, and feature importance estimations. This analysis facilitates a deeper understanding of the interactions between variables, assessing their influence on decision-making, and identifying potential shortcomings in decision-making. 
Following the analysis, modifications—such as redefining the reward function, reducing dimensionality, and assigning differentiated weights to input features—are implemented on the original \ac{RL} model to enhance its decision-making policy and overall network's performance.

\section{XRSIR}
This section exemplifies the use of eXplaNet to improve routing in \ac{SDN} networks. In particular, we enhanced RSIR, an \ac{RL}-based solution for routing in \ac{SDN} (see Figure 1(b)). The Control Plane gathers raw data on network status through periodic queries to the Data Plane. Such data is then processed by the Management Plane, where link status metrics are computed and stored. The \ac{RL} agent at the Knowledge Plane accesses such information to learn and determine the optimal routing paths. The Control Plane then retrieves the paths and installs them in the flow tables of the switches, preemptively setting up paths for incoming traffic. The \ac{RL} agent operates with a Q-learning model to continuously adjust the routing policies based on rewards that seek to minimize latency, optimize bandwidth, and reduce packet loss to maximize network performance dynamically.

\subsection{XRSIR}
Figure 1(b) introduces XRSIR, which was achieved by enforcing the pipeline outlined in Figure 1(a) into RSIR; XRSIR is an optimized version of RSIR. Data is collected from the Q-table generated by RSIR. The gathered data constitute the dataset used to train surrogate models for RSIR; the closest is selected. XAI techniques are employed to obtain local and global explanations of the RSIR's reward function behavior from the selected surrogate model. Such explanations allow network operators to understand how RSIR makes routing decisions and identify the features that exert the most significant influence on the reward function and, consequently, on the decisions taken. The reward function is tuned according to the explanations inferred to get XRSIR.

\subsubsection{Dataset Creation}
\final{After validating the implementation of the RSIR prototype by replicating the results reported in \cite{6rsir}, a dataset was generated to train the surrogate models. To ensure representativeness across diverse operating conditions, 16 distinct real intra-domain \ac{TM} were enforced on a network operating with the GÉANT (\ie the European National Research and Education Networks) topology, thereby capturing natural variations in network load as well as the different routing policies learned by the agent. The dataset was constructed by extracting samples that map the GEANT network state to the value of the agent’s actions. Each sample comprises: (i) performance metrics for each link (available bandwidth, delay, and packet loss) and (ii) information regarding the calculated routes. The Q-value, derived by the Q-learning agent through a reward function that integrates these metrics, served as the target variable.}

\final{During dataset construction, preprocessing was performed to ensure integrity and analytical relevance. Duplicate entries were removed, missing values imputed, and outliers in the target variable mitigated using an Inter-Quartile Range threshold. Categorical features representing origin and destination were target-encoded to capture their association with the target. These steps produced a consistent, tractable dataset, enhancing surrogate model validity and computational efficiency. The final dataset was stored in \ac{CSV} format file.
}

For the above-described dataset construction, in the Data plane, we used Mininet 2.3.0 with Open vSwitches 2.13.8 and a Python script to build a GÉANT topology with 23 nodes and 37 links. We scaled the original capacity of the links in Mininet from 10 Gbps, 2.5 Gbps, and 155 Mbps to 100 Mbps, 25 Mbps, and 1.55 Mbps~\cite{6rsir}. Each switch in the Data Plane had an assigned host responsible for forwarding and receiving \acp{TM} that were tagged by time, with peak hours due to higher traffic intensity. The traffic load was adjusted proportionally to the link's capacity. We also deployed the Ryu controller, Data Processing Module, Management Plane, Knowledge Plane, and the \ac{RL} agent provided by the RSIR prototype, which utilizes Python 2.7 with the Pandas 0.22 library. The Control, Management, and Knowledge planes were run on an Ubuntu 16.04 \ac{VM} with a Core i5-12500 processor and 8 GB of RAM. The \acp{VM} used for dataset creation were hosted on an Ubuntu Desktop 20.04 with an Intel Core i5-12500 processor and 16 GB of RAM. The \acp{VM} communicated via Transmission Control Protocol. The Control Plane and Data Plane communicated via OpenFlow 1.3.

\subsubsection{Surrogate Model Creation}
We used surrogate models to explain and enhance RSIR since the states of its \ac{RL} agent lack intrinsic characteristics. In RSIR, each state corresponds to a switch in the Data Plane represented as a graph, with edges symbolizing the links that interconnect these switches. The edges capture dynamic features such as available bandwidth, loss ratio, and delay, which the \ac{RL} agent uses to compute rewards that guide the routing between node pairs in the network. In this regard, surrogate models offer an efficient solution to reduce the RSIR's reward complexity.

\final{
The selection of the surrogate model considers the trade-off among fidelity, interpretability, and model complexity. Fidelity is the capacity to accurately replicate the behavior of the \ac{RL} agent, while interpretability facilitates the extraction of actionable insights from the model. The inverse relationship between interpretability and model complexity guided the selection. Given that the replication of the RSIR's reward function constitutes a regression problem, we considered four candidate models: linear regression, Ridge regression, Random Forest, and XGBoost, each trained on the generated dataset using Python 3.8.10 on Google Colab. The comparative evaluation was conducted regarding the coefficient of determination ($R^2$) and the \ac{MSE}, two widely adopted performance indicators in regression analysis. Among the tested approaches, XGBoost exhibited superior performance, achieving an $R^2$ of 0.91 and an \ac{MSE} of 225.39. In contrast, Random Forest attained an $R^2$ of 0.87 with an \ac{MSE} of 641.78, Ridge regression achieved 0.82 and 882.44, while linear regression reached 0.80 and 887.41, respectively. These results highlight the effectiveness of XGBoost in providing a high-fidelity surrogate while preserving computational efficiency.}

\final{Among the evaluated approaches, the more sophisticated ensemble methods, namely XGBoost and Random Forest, demonstrated the ability to capture the nonlinear dynamics of the reward function. Their feature importance scores aligned closely with the observed experimental outcomes, thereby identifying the variables most relevant for policy adjustment and subsequent system performance enhancement. In contrast, linear models failed to capture such complex relationships, leading to explanations that induced suboptimal policies and, consequently, degraded performance. Based on these findings, XGBoost was selected as the surrogate model, as its higher fidelity ensures that the derived explanations accurately reflect the RSIR agent’s behavior, thereby enabling more effective and reliable policy improvements.
}


\subsubsection{Explainability Techniques and Analysis}

\final{The application of feature relevance techniques allows for a quantitative assessment of each variable’s impact on the predictions generated by the XGBoost surrogate model. Insights derived from this analysis can subsequently inform refinements to the RSIR reward function, thereby improving overall system performance. The RSIR reward function is designed to depend directly on both delay ($delay$) and packet loss ($pkloss$), and inversely on the available bandwidth ($bwd$) of the links. Formally, the reward function is expressed as $R_{RSIR} = \beta_{1} \cdot \hat{bwd} + \beta_{2} \cdot \hat{delay} + \beta_{3} \cdot \hat{pkloss}.$ All metrics are normalized to prevent any single link state from disproportionately influencing the learning process of the \ac{RL} agent. The objective is to minimize the reward, thereby guiding the agent toward the selection of optimal network paths.}

\final{Feature relevance techniques, including \ac{SHAP}, \ac{PDP}, and \ac{ICE}, were applied to the XGBoost surrogate model to interpret its predictive behavior. \ac{SHAP} was implemented using the `shap` library with the TreeExplainer module, selected for its native compatibility with tree-based models such as XGBoost and its capacity to efficiently compute accurate feature contributions to individual predictions \cite{xai6g}. To complement both global and instance-level analyses, \ac{PDP} and \ac{ICE} plots were generated using the `PartialDependenceDisplay` function in the scikit-learn library. \acp{PDP} were employed to estimate the average marginal effect of each variable on the model's predictions, and \ac{ICE} profiles, obtained with the `kind= individual' parameter, were used to capture the variability of these effects across specific instances. This approach revealed nonlinear behaviors that may remain obscured in analyses based solely on average effects.
}

\begin{figure*}[!ht]
    \centering
    \includegraphics[width=0.66\linewidth]{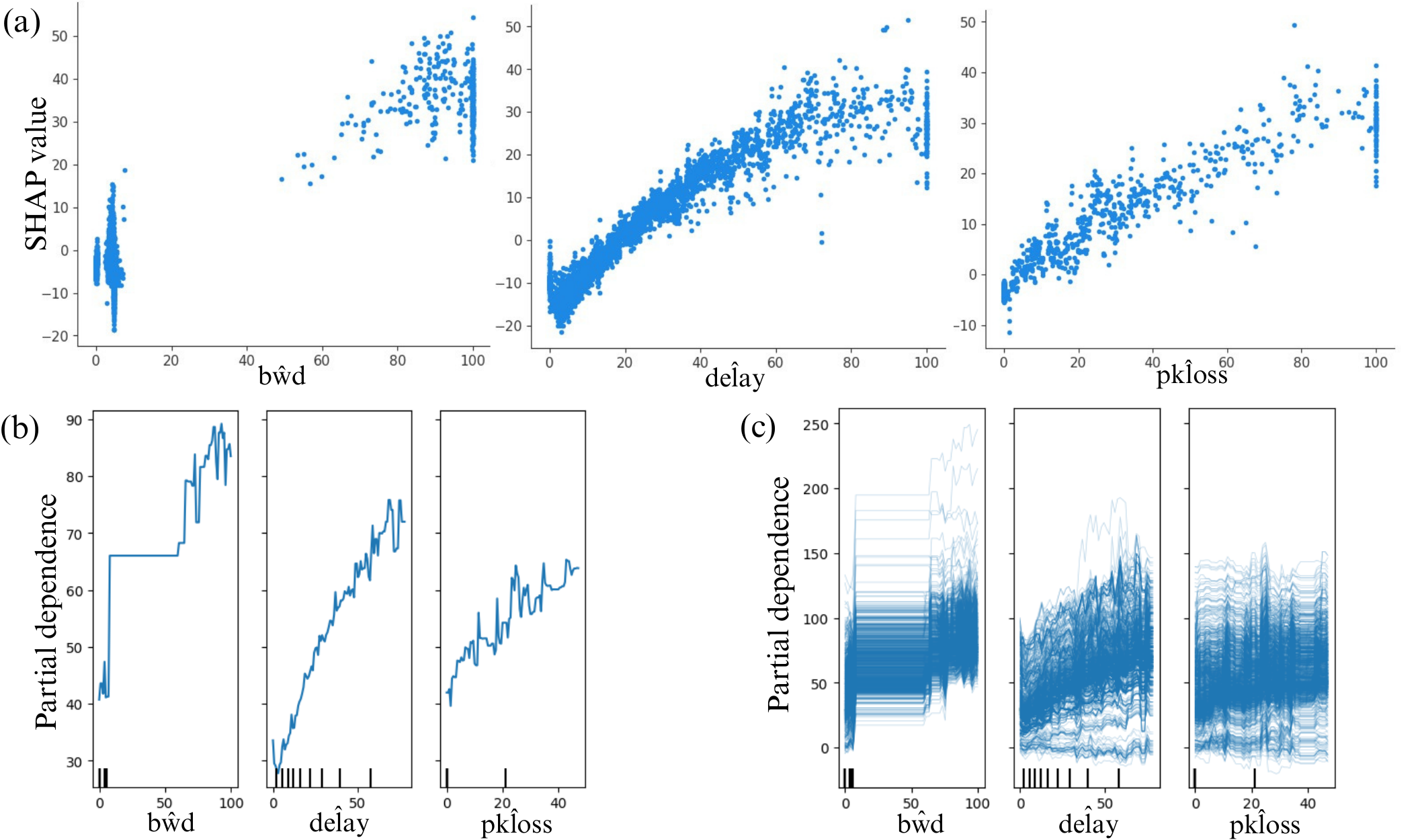}
     \vspace{-2mm}
    \caption{\final{Model interpretability for the XGBoost surrogate: (a) \ac{SHAP} dependence plots, (b) \ac{PDP}, and (c) \ac{ICE} illustrating the marginal and instance-level effects of the normalized features $\hat{bwd}$, $\hat{delay}$, and $\hat{pkloss}$ on the predicted performance.}}
    
    \label{fig:shap}
    \vspace{-3mm} 
\end{figure*}

Figure 2(a) shows a \ac{SHAP} dependence plot for the features under study in the surrogate model. The plots of $\hat{delay}$ and $\hat{pkloss}$ present a scatter in the data but show continuous growth, indicating that the Q-values increase when the values of such features rise, which is detrimental to the \ac{RL} agent's performance. The \ac{SHAP} plot of $\hat{bwd}$ exhibits a concentration of negative values when there is much available bandwidth, suggesting that this feature is the most relevant to minimizing the Q-values. This result indicates that RSIR prefers paths with the most available bandwidth. Figure 2(b) shows the dependency plot generated with \ac{PDP} for the surrogate model, highlighting how each feature globally influences the predictions. As in the \ac{SHAP} evaluation, $\hat{bwd}$ is the variable that impacts the \ac{RL} agent's performance the most; its increase leads to high Q-values, which is detrimental to the surrogate's performance. $\hat{delay}$ presents a continuous positive relationship, indicating that higher $\hat{delay}$ values increase Q-values. $\hat{pkloss}$ is the least significant feature since its slope is less steep than the other features and has a minor impact on increasing Q-values. Figure 2(c) depicts how features affect individual samples using the \ac{ICE} plots. Results reveal that a high $\hat{bwd}$ and a high $\hat{delay}$ impact Q-values, although the slope varies between observations. In $\hat{pkloss}$, a more heterogeneous response is observed, suggesting different sensitivities between the Q-values of samples and packet loss.

Results from applying XAI to analyze features' importance in RSIR's reward function, represented by the training dataset features of the surrogate model, indicated $\hat{bwd}$ and $\hat{delay}$ as the most influential variables, while $\hat{pkloss}$ showed to be the least one, although still of significant relevance. These findings underlined the need to adjust the RSIR's reward function to stand out $\hat{bwd}$ and $\hat{delay}$ and not point out to $\hat{pkloss}$.

\subsection{XRSIR's performance}
\final{The reward function of the \ac{RSIR} model, which assigned equal weights to all features, thereby neglecting their relative significance, was adjusted using insights derived from XAI techniques, resulting in an optimized variant termed XRSIR. Initially, through iterative experiments, feature weights were systematically varied in proportion to their observed importance to identify an empirical optimum $\beta$ for each feature. The experiments validated the guidance provided by XAI, demonstrating that prioritizing available bandwidth ($bwd$), followed by delay ($delay$) with balanced emphasis, and de-emphasizing packet loss ($pkloss$) enhances overall system performance. The resulting reward function, which is minimized to identify optimal paths, is expressed as $R_{XRSIR} = \beta_{1} \cdot \hat{bwd} + \beta_{2} \cdot \hat{delay} + \beta_{3} \cdot \hat{pkloss}.$}

\final{The two configurations exhibiting the highest overall performance were selected for further evaluation: XRSIR\textsubscript{1} (i.e., $\beta_{1}$ = 0.6, $\beta_{2}$ = 0.3, and $\beta_{3}$ = 0.1) and XRSIR\textsubscript{2} (i.e., $\beta_{1}$ = 0.65, $\beta_{2}$ = 0.35, and $\beta_{3}$ = 0). Their performance was assessed using four key metrics: average stretch, defined as the ratio between the length of the obtained route and the shortest possible route; average link delay; average link throughput; and average link loss rate. All metrics were computed over an identical time interval to ensure consistency and comparability of the results.
}

\begin{figure*}[ht!] 
\centering 
\subfloat
{\includegraphics[width=0.45\linewidth]{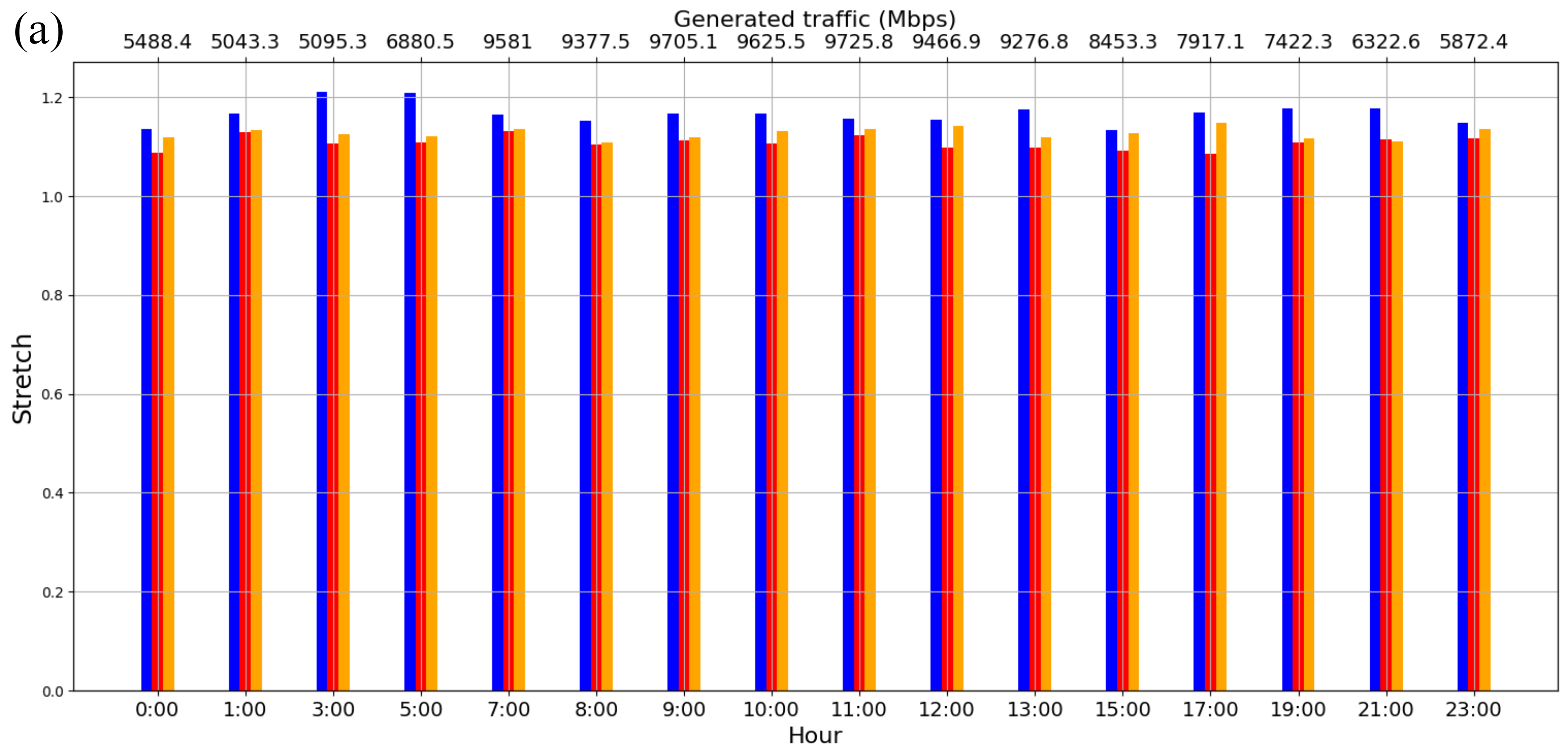} 
\label{fig:stretch}}\hfill 
\subfloat
{\includegraphics[width=0.45\linewidth]{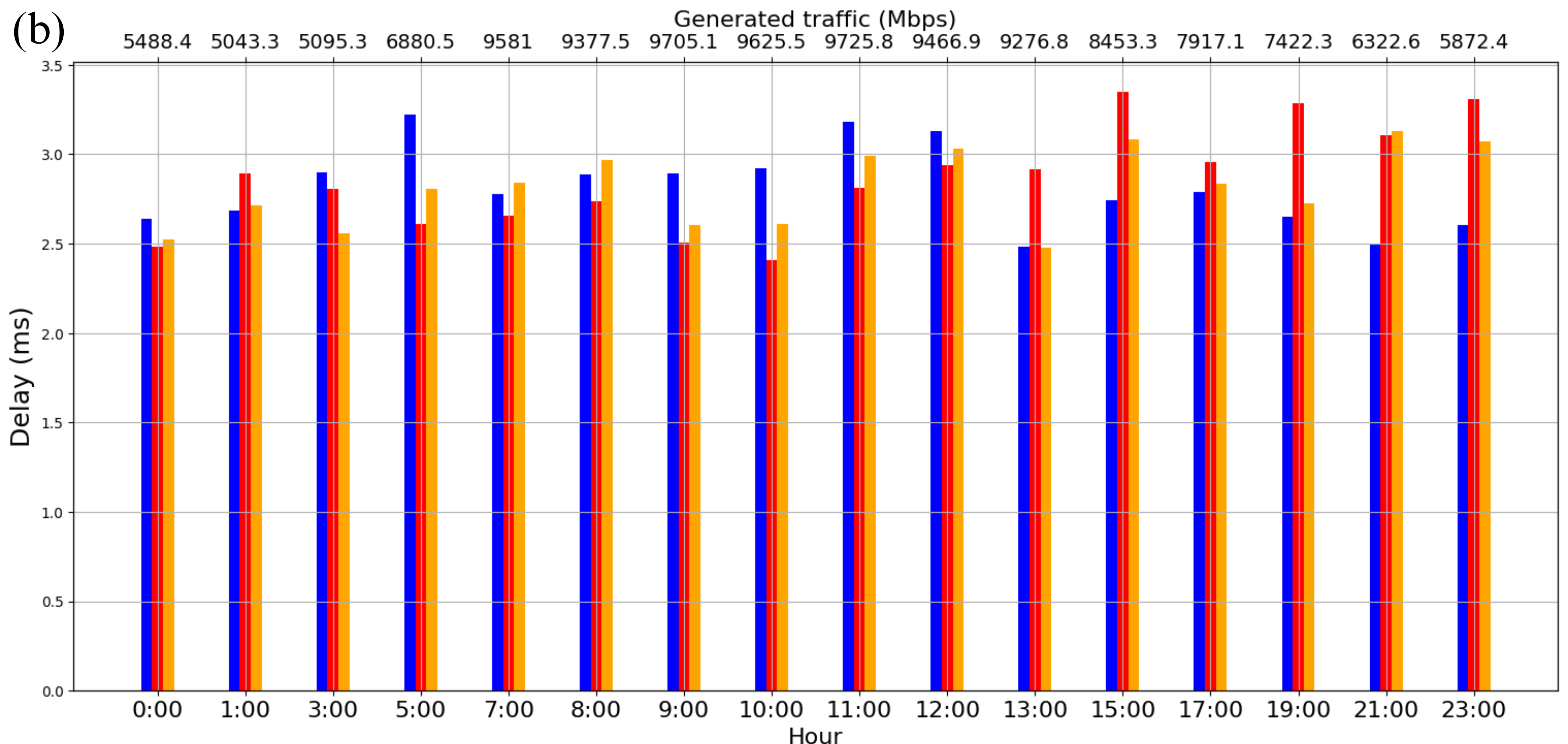}
\label{fig:delay}} \vspace{-3mm} 
\\ \subfloat
{\includegraphics[width=0.45\linewidth]{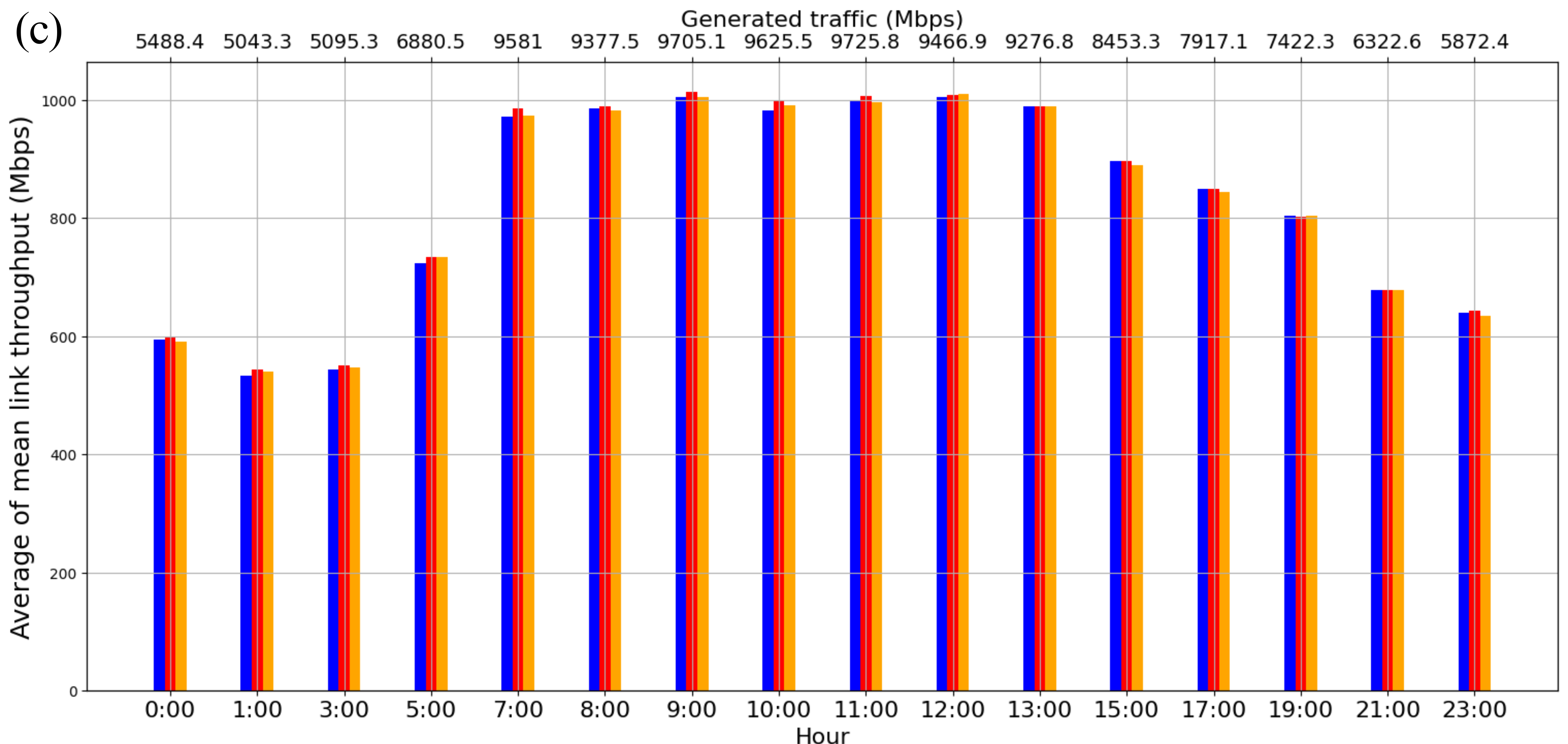} 
\label{fig:bwd}}\hfill 
\subfloat
{\includegraphics[width=0.45\linewidth]{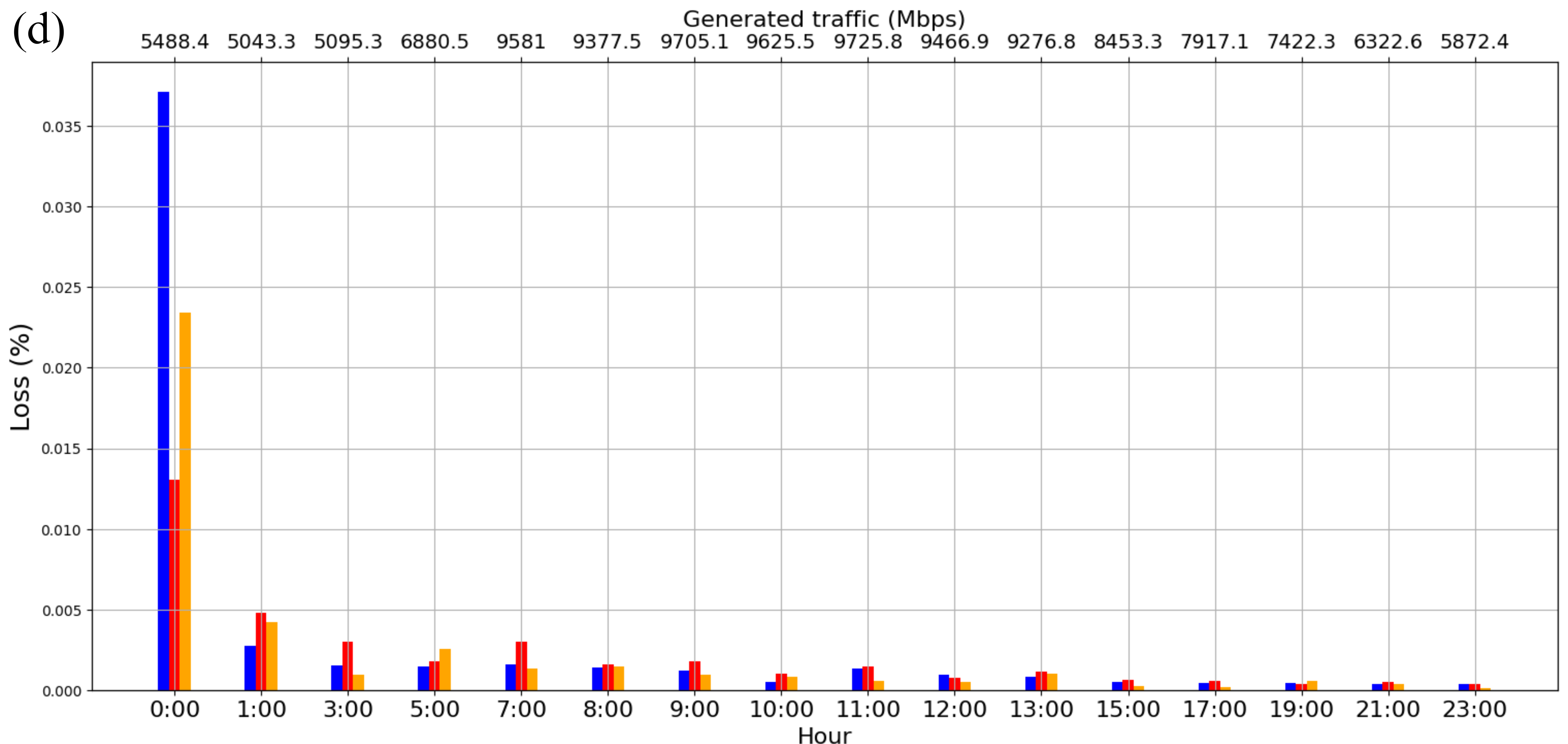} 
\label{fig:lossxrsir}}

\vspace{1mm} 
\includegraphics[width=0.35\linewidth]
{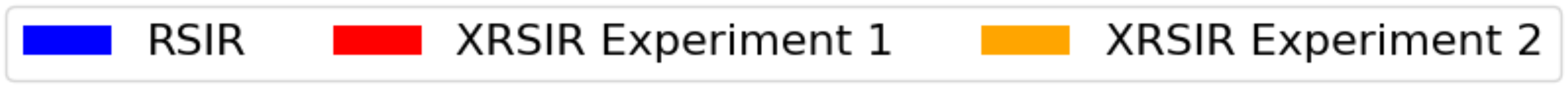} 
\caption{XRISR performance. (a) Mean stretch (b) Mean link delay (c) Mean link throughput (d) Mean link loss ratio} \label{fig:xrsirResults} \end{figure*}

Figures \ref{fig:stretch} and \ref{fig:delay} show the mean stretch and mean link delay obtained by RSIR, XRSIR\textsubscript{1}, and XRSIR\textsubscript{2}. Results reveal that for all \acp{TM}, XRSIR flavors selected shorter routes than RSIR. At best, XRSIR\textsubscript{1} (8.66\%) and XRSIR\textsubscript{2} (7.28\%) selected shorter routes than RSIR. These routes led to XRSIR\textsubscript{1} (19\%) and XRSIR\textsubscript{2} (13\%) getting lower mean link delay than RSIR in the \acp{TM} (from 5 am to 1 pm), which were passing the highest traffic. During low traffic, RSIR achieved a lower ($\eqsim$ 19\% at best) mean link delay than XRSIR\textsubscript{1} and XRSIR\textsubscript{2}.


  
Figures \ref{fig:bwd} and \ref{fig:lossxrsir} depict the mean link throughput and the mean link loss ratio along the day obtained by RSIR and the XRSIR flavors. XRSIR\textsubscript{1} (1.67\% at best) and XRSIR\textsubscript{2} (at highest 1.44\%)  achieved superior throughput than RSIR in 14 of the 16 and 7 of 16 analyzed \acp{TM}, respectively. In the other \acp{TM}, the difference regarding mean link throughput between RSIR and XRSIR is negligible. Regarding the mean link loss ratio, XRSIR\textsubscript{1} outperformed RSIR in 3 of 16 \acp{TM}, while XRSIR\textsubscript{2} outperformed during 9 out of 16. At best, during high traffic, XRSIR\textsubscript{1} (64\%) and XRSIR\textsubscript{2} (69\%) got lower loss than RSIR. Overall, XRSIR\textsubscript{2} reduced by 9\% the loss ratio.

\final{The results indicate that R\textsubscript{RSIR} is more suitable for low-traffic conditions, whereas R\textsubscript{XRSIR} is preferable under high-traffic scenarios. The XRSIR variants outperform RSIR in congested networks due to their reward function, which places greater emphasis on both $bwd$ and $delay$. In conditions of high congestion, link saturation constitutes the primary contributor to increased delay and packet loss. By prioritizing $bwd$, XRSIR effectively selects routes with higher capacity, thereby mitigating the adverse effects of network congestion.
}

\final{Under low-traffic conditions, $bwd$ no longer serves as the limiting factor. When employing balanced weights, RSIR is capable of identifying marginally shorter routes in terms of hop count, resulting in a modest reduction in total delay within lightly loaded networks. In contrast, XRSIR, by prioritizing $bwd$, may select routes with higher capacity but an additional hop, leading to a slight increase in delay under low-load conditions. These observations highlight the value of explanatory analyses for understanding agent decision-making and for the optimization of \ac{RL}–based network management strategies.
}

\section{Future Research Directions}

\ac{RL} models often are opaque systems, which
can hinder their broader adoption. Enhancing their practicality and
performance through XAI in
networking represents a compelling research opportunity. Key
directions are outlined below:

\final{\textbf{Adaptive reward functions.} A key limitation of the current approach is the static reward function, where weights ($\beta$) remain fixed. XAI analysis shows that the importance of $delay$ and $bwd$ changes with traffic and QoS objectives. We propose a real-time, automatic adjustment of weights based on network state and active application requirements, allowing the RL agent to better align decisions with dynamic network priorities.}

\final{\textbf{Explaining \ac{DRL} models.} The case study presented in this paper focuses on RSIR, which is implemented using a tabular Q-Learning algorithm. This approach exhibits limited scalability in large networks with extensive state and action spaces, which is commonly encountered in real-world applications. To address this challenge, we propose using eXplanet in DRL-based solutions, as well as the use of more complex surrogate models, such as \acp{DNN} and \acp{CNN}. Providing explanations for and refining \ac{DRL} agents—such as the one introduced in DRSIR \cite{9634122}—is critical for enhancing the transparency, reliability, and operational feasibility of \ac{DRL}-governed systems in large-scale network scenarios.}

\section{Closing Remark}
\label{sec:final-consideration}

RL algorithms often lack explainability and transparency, limiting their commercial viability in networking. XAI offers a solution to these challenges, enhancing RL deployment in real-world networks. This paper presents an XAI-driven pipeline for optimizing RL-based networking solutions, demonstrated through an SDN routing case study, and discusses the broader opportunities and challenges of integrating XAI to improve trust, usability, and performance in intelligent networks.



\bibliography{reference}
\bibliographystyle{IEEEtran}
\vspace{-1cm}
\begin{IEEEbiographynophoto}{Yeison S. Murcia} is pursuing his M.Sc. degree  at the State University of Campinas,  Brazil.
\end{IEEEbiographynophoto}
\vspace{-1.30cm}
\begin{IEEEbiographynophoto}{Oscar M. Caicedo} [S’11, M’15, SM’20] is a full professor at the
Universidad del Cauca, Colombia.
\end{IEEEbiographynophoto}
\vspace{-1.3cm}
\begin{IEEEbiographynophoto}{Daniela M. Casas} [S15] is pursuing her Ph.D. degree  at the University of Campinas,  Brazil.
\end{IEEEbiographynophoto}
\vspace{-1.3cm}
\begin{IEEEbiographynophoto}{Nelson L. S. da Fonseca}  [M’88, SM’01] is a Full Professor at the
Institute of Computing, University of Campinas, Brazil. He serves as the IEEE ComSoc President-Elect. 

\end{IEEEbiographynophoto}
\vspace{-1cm}
\end{document}